# Synthesis and study of (Na, Zr) and (Ca, Zr) phosphate-molybdates and phosphate-tungstates: Thermal expansion behavior, radiation test and hydrolytic stability

M.E. Karaeva[1], D.O. Savinykh[1], A.I. Orlova[1], S.A. Khainakov[2], A.V. Nokhrin[1(*)], M.S. Boldin[1], S. Garcia-Granda[2], A.A. Murashov[1], V.N. Chuvil'deev[1], V.A. Skuratov[3,4,5], N.S. Kirilkin[3], P.Yu. Yunin[1,6], and N.Yu. Tabachkova[7,8]

[1] Lobachevsky State University of Nizhny Novgorod, 603022, Nizhniy Novgorod, Russia

[2] Oviedo University-CINN, 33006, Oviedo, Spain

[3] Joint Institute of Nuclear Research, 141980, Dubna, Russia

[4] National Research Nuclear University MEPhI (Moscow Engineering Physics Institute), 115409, Moscow, Russia

[5] Dubna State University, 181982, Dubna, Russia

[6] Institute for Physics of Microstructure, Russian Academy of Science, 603950, Nizhniy Novgorod, Russia

[7] Prokhorov General Physics Institute, Russian Academy of Science, 119991, Moscow, Russia

[8] National University of Science and Technology "MISIS", 119049, Moscow, Russia

e-mail: albina.orlova@gmail.com

**Abstract**

Thermal expansion behavior at high temperatures of synthesized $Na_{1-x}Zr_2(PO_4)_{3-x}(XO_4)_x$, $Ca_{1-x}Zr_2(PO_4)_{3-x}(XO_4)_x$, X = Mo, W ($0 \leq x \leq 0.5$) compounds has been investigated. Ceramics with relatively high density (more than 97.5%) were produced by Spark Plasma Sintering (SPS) of submicron powders obtained by sol-gel synthesis. The study of strength characteristics has revealed that hardness the ceramics are greater than 5 GPa, and minimum fracture toughness factor was 1

---

(*) Corresponding author (23/3 Gagarina ave., Nizhniy Novgorod, 603022, Russian Federation, nokhrin@nifti.unn.ru)

MPa·m$^{1/2}$. It was found that ceramics have a high hydrolytic resistance in the static regime - the minimum leaching rates for the Mo- and W-containing specimens were 31·10$^{-6}$ and 3.36·10$^{-6}$ g/(cm$^2$·day), respectively. The ceramics had a high resistance to the irradiation by Xe$^{+26}$ multiple-charged ions with the energy 167 MeV up to the fluences in the range 1·10$^{12}$ – 6·10$^{13}$ cm$^{-2}$. The Mo-containing Na$_{0.5}$Zr$_2$(PO$_4$)$_{2.5}$(XO$_4$)$_{0.5}$ ceramics were shown to have a higher radiation resistance that the phosphate-tungstates.

### 1. Introduction

At present, the inorganic compounds with the NaZr$_2$(PO$_4$)$_3$ (NAISICON, NZP)-type structure are being studied extensively in connection with the prospects of application of these ones for solving the problems of immobilization of the highly active components of the radioactive waste [1-3]. In particular, the Mo-containing fractions of the radioactive waste solidify as the inclusions in glass together with other nuclides, for example, the wedge-shaped glass ceramics in the borosilicate glasses [4-6]. When more than 10 wt.% of Mo is incorporate into a multi-component highly radioactive waste, Mo can form powellite – a mineral of Ca molybdate, CaMoO$_4$. This affects the chemical resistance of the glasses adversely affecting the solubility of molybdates [6]. Similar situation takes place with W.

Stable crystalline mineral-like matrix substances are most suitable for binding Mo and W. The inclusion of Mo in NASICON-like ceramics leads to a decrease in the Mo leaching rate as compared to the glasses, glass ceramics, and Synroc materials containing individual phases of readily soluble molybdates [7]. The sintering of the crystalline compounds into ceramics increases the relative density of the ones preserving the phase composition. At the same time, the resulting ceramics have high strength characteristics [1, 8-11].

The compounds with the NaZr$_2$(PO$_4$)$_3$ structure are characterized by the crystal chemical formula (M1)$^{VI}$(M2)$^{VIII}_3$[L$^{VI}_2$(XO$_4$)$_3$]$^{n-}$ where M1 and M2 are the positions in the voids of the [LV1$_2$(XO$_4$)$_3$]$^{n-}$ framework. The family of phosphates with the NZP structure includes many compounds due to the possibility of isomorphic substitutions in various positions of the structure [12-14]. The framework of the structure is formed by multiply charged cations L with an oxidation states 5+, 4+, 3+, or 2+ of small sizes and by anions XO$^{4-}$. Most compounds of the NASICON family contain P as an anion-forming element X. However, there are also compounds with the NASICON structure, which P is replaced by other anions in. Some compounds are known where P is replaced by Si [8, 15-18], by S [19, 20], by V [21], by As [22], by Se [23], and by Mo [7, 24, 25].

The positions of the M- type can be occupied fully or partially or remain vacant. The composition of NASICON phosphates can include cations in the oxidation states from +1 to +4; mainly, the population with low-charged and relatively large cations occurs. The presence of four positions (M1, M2, L, and X) capable of incorporating cations of various sizes determines the prospects for application of materials based on compounds with the NASICON structure in various fields. The compounds of this structure are feature by high ionic conductivity as well as by high corrosion, thermal, radiation, and chemical resistance along with high catalytic activity [25, 26].

The behavior of phosphates under heating and the values of thermal expansion coefficients depend on the nature of the constituent ions, the charge, size, and electronegativity of the ones. Due to the ability of the structure to accommodate various constituent elements, it becomes possible to develop various materials with desired thermal expansion coefficients [27]. In most cases, these compounds are featured by expansion of the unit cell along the *c* crystallographic axis and by compression along the *a* and *b* axes when heated (anisotropic thermal expansion). In the case of a partial or complete substitution of the anions, the charge (n) of the framework changes. In this case, the cations in the M-sites compensate this charge preserving the electro-neutrality. Therefore, the cations in the M sites and the total occupancy of the sites affect the changes in the thermal expansion coefficients. Some of these compounds have small or ultra-low [down to (1-2)·10$^{-6}$ deg$^{-1}$] or controlled thermal expansion coefficients [1, 28-35]. Also, these ones are stable under hydrothermal conditions at the temperatures up to 400 °C and at the durations of contact with water up to two years [1, 36-39]. Such a unique hydrolytic resistance of the ceramics with the NASICON structure increases an interest to these ones as the materials for solving the radiochemical problems of immobilization of radioactive waste [1, 12].

At present, hot pressing or free sintering of preliminary pressed powders are applied for sintering the mineral-like ceramics most often [9-11, 40]. A necessity of long-time holding at elevated sintering temperatures to provide a high density of the ceramics is the drawback of these methods. Spark Plasma Sintering (SPS) stands out of other methods of obtaining the ceramic samples. The method consists in a high-speed heating (up to 2500 °C/min) of a powder material in vacuum by passing a series of short electric current pulses through a sample and a mold while applying a pressure [41-43]. This method is featured by high shrinkage rates at reduced sintering temperatures. The sintering of the mineral-like ceramic samples up to high relative densities (about 95%) takes place in shorter times [18, 44-46]. This ensures the reduction of the times of handling the highly active components of the radioactive waste as compared to traditional hot pressing methods and free sintering of the powders pressed in advance. These advantages of SPS cause an intensive expansion of applications in the development of the ceramic matrices for the immobilization of high-level waste (HLW) and in transmutation of minor actinides [18, 44-49].

Ceramic materials obtained by SPS are featured by a high relative density and enhanced physical and mechanical properties, which open up new possibilities for obtaining the ceramic materials for various purposes [18, 41-45]. Some of these ceramics have a high radiation resistance [45, 46, 50] and high hydrolytic resistance [49, 51, 52].

This work is devoted to the synthesis and investigation of properties of the NASICON-framework orthophosphates with partial substitution of the P ions by Mo and W and various cations in the structure voids. These ceramics may be applied potentially for the immobilization of the Mo- and W-containing fractions of radioactive waste.

## 2. Materials and methods

Solid solutions of the $Na_{1-x}Zr_2(PO_4)_{3-x}(XO_4)_x$ and $Ca_{1-x}Zr_2(PO_4)_{3-x}(XO_4)_x$, types where X = Mo, W, and x = 0.1, 0.2, 0.3, 0.4, 0.5 have been chosen as the objects of investigations. The compounds were synthesized by sol-gel method. A sample of Zr oxychloride was dissolved in a mixture of solutions of ammonium molybdate or tungstate and Na or Ca nitrate. A 1M solution of ammonium dihydrogen phosphate was added dropwise to the resulting solution in the course of continuous stirring. As a result, gel-like precipitates were formed. The reagents were mixed in the stoichiometric proportions. For more complete precipitation, a salting-out agent (ethyl alcohol) was added. After short-time stirring, the gel was dried at 90 °C for 1 day. The resulting powders were annealed in two stages at 600 and 800 °C for 20 h at each stage with the dispersion in an agate mortar and investigation by X-ray diffraction (XRD) between the stages.

The ceramics were sintered from the obtained powders by SPS using Dr. Sinter™ model SPS-625 (SPS SYNTEX®, Japan). Prior to sintering, the powders were ground up in an agate mortar during 5 min. The powders were placed in a graphite mold with an inner diameter of 12.8 mm and heated up by passing the millisecond pulses of high-power direct electric current. The temperature was measured with a Chino IR-AHS2 pyrometer (Chino Corporation, Japan) focused onto the graphite mold surface. Sintering was performed in vacuum. A two-stage heating regime was used - stage A: heating up to 570-580 °C with the heating rate 100 °C/min, stage B: heating up to the sintering temperature $T_s$ with the heating rate 50 °C/min. The holding at the sintering temperature $T_s$ was absent ($t_s$ = 0 s). The average uniaxial pressure applied was ~65 MPa. In order to avoid the cracking of the ceramic specimen, a multistage pressure regulation during sintering was applied: (stage I) a gradual increasing of the uniaxial pressure P from 35-38 MPa up to 64-65 MPa during the rapid heating stage up to 570-580 °C, (stage II) holding at 64-66 MPa at 580-840 °C; (stage III) reducing the pressure down to 60 MPa at 900 °C, and (stage IV) gradual increasing the pressure again up to ~65 MPa to the moment of the end of shrinkage. Typical scheme of the pressure and temperature regulations during SPS is presented in Fig. 1. The samples were cooled down together with the setup (stage C, Fig. 1). As a result of sintering, dense ceramic specimens without macrodefects were obtained.

The temperature dependencies of the shrinkage L and of the shrinkage rate S were measured in the course of sintering using a built-in dilatometer of SPS-625 setup.

The XRD phase analysis of the powders and ceramics was performed using Bruker® D8 Discover™ X-ray diffractometer in symmetric Bragg-Brentano geometry. An X-ray tube with a Cu cathode (CuK$_\alpha$ radiation) was used. The goniometer radius was 30 cm. All experiments were performed in the same conditions. The specimens comprising cylindrical tablets of 12 mm in diameter were placed on the goniometer table. At the beginning of each experiment, a procedure of adjustment of the specimen height by beam splitting method was performed with 0.1 mm slits both on the primary beam and in front of the detector. When acquiring the XRD curves, the primary beam size was limited in the equator plane by a 0.6 mm slit, in the axial plane – by a 12 mm one. The XRD curves were acquired in the θ/2θ-scan mode in the angle range from 10° to 70° in 2θ. The step in angle was 0.1°. LynxEYE linear position-sensitive detector with 192 independent acquisition channels and 2° angle aperture in 2θ was used. The dwell time was 2 seconds that was equivalent to the effective dwell time about 40 seconds due to the use of the position-sensitive detector. The thermal expansion in the temperature range 25-800 ºC was investigated using Panalytical® X'Pert Pro™ high temperature X-ray diffractometer (Malven PANAlytical B.V., Netherlands) with Anton Paar® HTK-1200N high-temperature chamber (Anton Paar GmnH, Austria) with a step of 100 ºC. The unit cell parameters were calculated by Rietveld method using Topas software.

The microstructure of the powders and ceramics was investigated using Jeol® JSM-6490 scanning electron microscope (SEM, Jeol Ltd., Japan) with Oxford Instruments® INCA 350 energy dispersion (EDS) microanalyzer (Oxford Instruments pls., UK) and Jeol JEM-2100 transmission electron microscope (TEM, Jeol Ltd., Japan). The mean particle size (D) and grain size (d) was calculated by section method using GoodGrains 2.0 software (UNN, Russia).

The density of the obtained ceramics was measured by hydrostatic weighing in distilled water using Sartorius® CPA 225D balance (Sartorius AG, Germany). The uncertainty of the density measurement was ± 0.001 g/cm$^3$. The theoretical density of the ceramics ($\rho_{th}$) was calculated on the base of the XRD investigations. The microhardness ($H_v$) of the ceramics was measured using Duramin® Struers-5 hardness tester (Struers Inc., USA). The load was 200 g. The minimum fracture toughness factor ($K_{IC}$) was calculated according to the Palmquist method from the length of the largest radial crack forming by the Vickers pyramid during the indentation of the ceramics.

The investigations of the hydrolytic stability of the ceramic samples was carried out under static conditions according to Russian National Standard GOST R 52126-2003 "Radioactive waste. Determination of chemical resistance". The tests were performed in distilled water at room temperature (25-28 °C). The samplings of the contact solution were taken 1, 3, 7, 10, 14, 21, and 28 days after the start of the tests. The analysis of the solution samples for the content of Mo and W was performed by inductively coupled plasma mass spectrometry using ELEMENT™ 2 high resolution mass spectrometer (Thermo Scintific®, Bermen, Germany) with an external calibration. The calibration was performed using inductively-coupled plasma mass spectrometry ICP-MS-68A-B (High Purify Standards®, USA) using Thermo Scientific® ELEMENT™ 2 high-resolution mass spectrometer (Thermo Scientific, Germany) within the framework of an external calibration.

Radiation stability of ceramics was evaluated using high energy (167 MeV) Xe$^{+26}$ ion irradiation at IC-100 FLNR JINR cyclotron (Joint Institute of Nuclear Research, Russia). The specimens were irradiated at room temperature (23-27 °C) with fluences ranged from $1\cdot10^{12} - 6\cdot10^{13}$ cm$^{-2}$. Average ion flux was about $2\times10^9$ cm$^{-2}\cdot$s$^{-1}$ to avoid any significant heating of targets. The temperature of targets during irradiation did not exceed 30 °C. The uniform distribution of the ion beam over the irradiated target surface was achieved ion beam scanning. The accuracy of the ion flux and fluence measurements was 15%.

## 3. Results and discussion

### 3.1 Synthesis and studies of the powders

The powder samples were light-colored. According to the SEM results, the submicron-grained powders were agglomerated – there were some large agglomerated up to 10-20 μm in sizes in the $Ca_{1-x}Zr_2(PO_4)_{3-x}(XO_4)_x$ powders (Fig. 2a) and ~50 μm in size in the $Na_{1-x}Zr_2(PO_4)_{3-x}(XO_4)_x$ ones (Fig. 2c, e) after grinding in the agate mortar. No considerable effect of the W and Mo contents on the granulometric composition and morphology of the powders was observed.

Figs. 3 and 4 present TEM images of the $Ca_{1-x}Zr_2(PO_4)_{3-x}(MoO_4)_x$ powders with different Mo contents (Fig. 3) and of $Na_{0.5}Zr_2(PO_4)_{2.5}(XO_4)_{0.5}$ ones, X = Mo, W (Fig. 4). The $Ca_{1-x}Zr_2(PO_4)_{3-x}(MoO_4)_x$ powders were large-grained and agglomerated strongly. The individual particle sizes scattered from ~50-100 nm to ~300 nm and didn't depend on the Mo content (Fig. 3). The $Na_{0.5}Zr_2(PO_4)_{2.5}(XO_4)_{0.5}$ powders were more fine-dispersed and agglomerated. The particle sizes scattered from ~20-30 nm to ~50-100 nm (Fig. 4). The agglomerate sizes were ~0.2-0.3 μm. All powders had crystalline structure (Figs. 3b, d, f and 4b, d, e, f).

According to the XRD phase analysis data (Figs. 5 and 6), the compounds under study crystallized in the $NaZr_2(PO_4)_3$ type structure, hexagonal syngony, space group $R\bar{3}c$ (analog of $NaZr_2(PO_4)_3$ [29]) for the Na-containing samples and space group $R\bar{3}$ (analog of $Ca_{0.5}NaZr_2(PO_4)_3$ [53]) for the Ca-containing ones. The XRD patterns of the samples with Ca at x > 0.2 for phosphate-molybdates and x > 0.3 for phosphate-tungstates contained a significant number of auxiliary phase reflections. For this reason, these samples were not studied further.

The particle (coherent scattering region, CSR) sizes $D_{XRD}$ were estimated using Scherer equation: $D_{XRD} = K \cdot \lambda / \beta \cdot \cos\theta_{max}$ where $K = 0.9$ is the Scherer constant, $\lambda$ is the wavelength of the X-ray radiation, $\beta$ is the full width at half maximum of the XRD peak (in radians), and $\theta_{max}$ is the diffraction angle corresponding to the (116) XRD peak maximum. The analysis of the XRD results demonstrated the particle sizes $D_{XRD}$ of the $Na_{1-x}Zr_2(PO_4)_{3-x}(MoO_4)_x$ powders to decrease monotonously from 61 nm down to 19 nm with increasing Mo content from 0.1 up to 0.5 and from 44 nm down to 23 nm with increasing W content from 0.1 up to 0.5 (Fig. 7). No considerable effects of W and Mo contends on $D_{XRD}$ of the $Ca_{1-x}Zr_2(PO_4)_{3-x}(XO_4)_x$ powders were observed; the mean particle sizes were ~ 60 nm for the W-containing powders and ~ 47 nm for the $Ca_{1-x}Zr_2(PO_4)_{3-x}(MoO_4)_x$ ones. The analysis of the XRD, TEM and SEM results has allows concluding partial sintering of several nanoparticles into the submicron ones to take place in the course of synthesis, at the last stage of high-temperature annealing (800 °C, 20 hrs).

The unit cell parameters for the solid solutions were determined from the XRD data (the results are presented in Table 1).

The composition dependences of the unit cell parameters for the phosphate-molybdates and phosphate-tungstates are presented in Fig. 8. Substitution of phosphate anion $(PO_4)^{3-}$ ($R(P^{5+}) = 0.17$ Å) by larger molybdate anions $(MoO_4)^{2-}$ ($R(Mo^{6+}) = 0.41$ Å) and tungstate ones $(WO_4)^{2-}$ ($R(W^{6+}) = 0.44$ Å) leads to an increasing of the unit cell parameters. One can see from Fig. 8 the increasing of the W contents in the $Na_{1-x}Zr_2(PO_4)_{3-x}(XO_4)_x$ compounds to result in more essential increasing of the lattice parameter $c$ and of the unit cell volume $V$ than the increasing of the Mo one. No essential differences in the values of lattice parameter $a$ for the phosphate-molybdates and phosphate-tungstates $Na_{1-x}Zr_2(PO_4)_{3-x}(XO_4)_x$ were observed. for the Ca-phosphates-tungstates $Ca_{1-x}Zr_2(PO_4)_{3-x}(WO_4)_x$.

No laws of dependencies of the unit cell parameters $a$ and $c$ as well of the unit cell volume $V$ on the W contents expressed clearly were found. In the case of $Ca_{1-x}Zr_2(PO_4)_{3-x}(MoO_4)_x$, the increasing of the Mo content up to x = 0.2 resulted in a linear increasing of all unit cell parameters and, hence, in a linear increasing of the unit cell volume $V$.

To study the behavior of the obtained compounds upon heating, the XRD patterns of the samples were recorded at elevated temperatures (25-800 °C). The XRD data were used to calculate the values of the unit cell parameters at different temperatures. The temperature dependencies of the cell parameters are shown in Figs. 9 and 10. One can see from Fig. 9 an insufficient decreasing (~0.02 Å) of the unit cell parameter *a* and considerable increasing of the parameter *c* (~0.2-0.4 Å) with increasing temperature (from 20 up to 800 ºC) to be observed for $Na_{1-x}Zr_2(PO_4)_{3-x}(XO_4)_x$. It leads to the increasing of the unit cell volume *V* when heating. For $Ca_{1-x}Zr_2(PO_4)_{3-x}(XO_4)_x$, more essential increasing of the parameter *a* was observed at similar increasing of temperature (Fig. 10) that resulted in a weaker increasing of the unit cell volume of $Ca_{1-x}Zr_2(PO_4)_{3-x}(XO_4)_x$ as compared to $Na_{1-x}Zr_2(PO_4)_{3-x}(XO_4)_x$.

The above dependencies were used to calculate the values of the axial ($\alpha_a$ and $\alpha_c$), the average ($\alpha_{av}$), and the volume ($\beta$) thermal expansion coefficients (CTEs) as well as of the thermal expansion anisotropy ($\Delta\alpha$) for the phosphate-molybdates and phosphate-tungstates under study (Table 2, Figs. 11 and 12). The analysis of these data shows the decreasing of $\alpha_a$ and $\alpha_c$ with increasing W and Mo contents to be observed in $Na_{1-x}Zr_2(PO_4)_{3-x}(XO_4)_x$. At the same time, the mean value of $\alpha_{av}$ changes weakly. The volume CTE ($\beta$) decrease insufficiently whereas the thermal expansion anisotropy ($\Delta\alpha$) increases. In general, similar character of the CTE dependencies on the composition was observed for the $Ca_{1-x}Zr_2(PO_4)_{3-x}(MoO_4)_x$ compounds at the increasing of the Mo content (Fig. 11). For the W-containing phosphates $Ca_{1-x}Zr_2(PO_4)_{3-x}(WO_4)_x$, no considerable effect of the W content on all thermal expansion parameters ($\alpha_a$, $\alpha_c$, $\alpha_{av}$, $\beta$, and $\Delta\alpha$) was observed.

One can note a trend for the thermal expansion parameters to approach zero with decreasing population of the extraframe positions in the structure of the phosphate-molybdates and phosphate-tungstates under study. So far, novel ceramic materials with zero CTE and high thermal shock resistance can be developed potentially on base of the phosphates with predefined W and/or Mo contents studied in the present work.

### 3.2 Sintering and studies of the ceramics

The ceramic samples with high relative densities were obtained from the Na-containing compounds by SPS. The ceramic samples haven't visible macro- and microcracks. Examples of sintering diagrams L(T) and S(T) are shown in Fig. 13. The shapes of the dependencies L(T) are typical for SPS of the nano- and submicron NZP-type powders [18, 44, 46].

The analysis of the temperature dependencies of the shrinkage L(T) for the phosphate-molybdates (Fig. 13a, c) shows an increasing of the maximum shrinkage $L_{max}$ и and of the shrinkage rate $S_{max}$. with increasing Mo content from x = 0.1 up to 0.3 to be observed. At the same time, an increasing of the optimal sintering temperature was observed, which is manifested as the shift of the dependencies L(T) and S(T) towards higher heating temperatures. At further increasing of the Mo contents up to x = 0.5, a shift of the dependencies L(T) and S(T) towards lower sintering temperatures was observed. In our opinion, such a behavior is due, first of all, to the appearance of some auxiliary phases during sintering and to increasing contents of the ones with increasing Mo concentration (see above).

In the $Na_{1-x}Zr_2(PO_4)_{3-x}(WO_4)_x$ phosphates, an insufficient increasing of the shrinkage and decreasing of the optimal sintering temperature corresponding to the finish of the intensive shrinkage stage with increasing W content were observed (Fig. 13b). At the same time, an increasing of the maximum shrinkage rate of nanopowders $S_{max}$ more than 3 times and insufficient shift of the temperatures, which the maximum shrinkage rates $S_{max}$ were manifested at towards higher heating temperatures were observed (Fig. 13d).

The averaged SPS times were 13 min for the phosphate-molybdates $Na_{1-x}Zr_2(PO_4)_{3-x}(MoO_4)_x$ and 16 min for the phosphate-tungstates $Na_{1-x}Zr_2(PO_4)_{3-x}(WO_4)_x$. The characteristics of the sintering process (the heating rates $V_h$, the sintering temperatures $T_s$, and the full SPS process time $t_{sps}$ including the heating times), the relative densities achieved, the microhardness values, and the minimal fracture toughness factors for the ceramics obtained are presented in Table 3. One can see from Table 3 that the increasing of the W and Mo contents resulted in an increasing of the relative density of the $Na_{1-x}Zr_2(PO_4)_{3-x}(XO_4)_x$ ceramics. The densities close to the theoretical one were ensured for almost all ceramics. In the ceramics with increased W and Mo contents, the relative densities exceeded 100% insufficiently. In our opinion, this result originates from the auxiliary phases contained in the powders synthesized (see the XRD phase analysis results above).

Figures 14 and 15 present the SEM images of the microstructure of the sintered phosphate-molybdates $Na_{1-x}Zr_2(PO_4)_{3-x}(MoO_4)_x$ (Fig. 14) and phosphate-tungstates $Na_{1-x}Zr_2(PO_4)_{3-x}(WO_4)_x$ (Fig. 15) with various Mo and W contents. One can see in Fig. 14 a decreasing of the mean grain size (Fig. 14b, d) and an increasing of the microstructure uniformity of the ceramic phosphate-molybdates (Fig. 14a, c) with increasing Mo content to be observed. Note that there were some areas with abnormally large grains in the sintered ceramics with low Mo contents ($x = 0.1$). The sizes of these grains were ~5-10 μm (Fig. 14a) that were greater than the mean grain size of the ceramic matrix (Fig. 14b) by an order of magnitude. In the ceramics with increased Mo contents ($x = 0.5$), the areas with abnormally large grains were almost absent (Fig. 14c) whereas the mean grain sizes in the sintered ceramic phosphate-molybdates were close to 100-200 nm (Fig. 14d).

Similar effect has been observed when analyzing the results of SEM investigations of the ceramic phosphate-tungstates with different W contents. As one can see in Fig. 15, the increasing of the W content resulted in a decreasing of the mean grain sizes in the ceramics down to ~100-150 nm. There are few areas with abnormally large grains, the sizes of which is not greater than 2-5 μm in the ceramics with $x = 0.5$ (Fig. 15d).

The XRD phase analysis evidenced the phase composition of the ceramics not to change after sintering. An insufficient increasing of the intensities and decreasing of the widths of the XRD peaks have been observed that is typical for sintering the nano- and submicron powders, which is associated with the grain growth.

As one can see from Table 3, the increasing of the Mo and W contents resulted in an insufficient increasing of the microhardness of the ceramics up to 5.3-5.6 GPa. At the same time, the minimum fracture toughness factor almost didn't changed and exceeded 1 MPa·m$^{1/2}$ as a rule that is typical value for the mineral-like ceramics with the NZP structure [49].

The hydrolytic stability of the samples with high Mo and W contents (x = 0.4 and 0.5) has been studied. According to the XRD data, no crystal structure damage during the hydrolytic tests was observed. The unit cell parameters of $Na_{1-x}Zr_2(PO_4)_{3-x}(XO_4)_x$ were the same before and after the hydrolytic tests. The release rates per unit surface area ($R$) for particular components were determined according to the formulas:

$$R = NL/t, \qquad (1)$$

$$NL = m/(\omega \cdot S), \qquad (2)$$

where $m$ [g] is the mass of the component leached in given time period, $t$ [days] is the test duration, $S$ [cm$^2$] is the open surface area, and ω is the mass concentration of the component in the original sample.

The values of the normalized weight loss NL and of the leaching rates R obtained are presented in Table 4. The time dependencies of the above quantities are shown in Fig. 16. The minimum leaching rates $R_{min}$ achieved were $31.6 \cdot 10^{-6}$ g$^{-1} \cdot$cm$^{-2} \cdot$day$^{-1}$ for the Mo-containing compounds and $3.36 \cdot 10^{-6}$ g$^{-1} \cdot$cm$^{-2} \cdot$day$^{-1}$ for the W-containing ones. So far, the minimum leaching rate of Mo from the $Na_{1-x}Zr_2(PO_4)_{3-x}(XO_4)_x$ ceramics at x = 0.4 and 0.5 was greater than the one of W by an order of magnitude[1]. This is a very low leaching rate for the ceramics with the NZP structure evidencing a high chemical resistance of these ones. In our opinion, this result evidences the inorganic compounds of the NZP family to be promising for applications as the binders for the W- and Mo-containing fractions of the radioactive waste.

---

[1] As follows from Table 4 and Fig. 16, the leaching rate R decreases with increasing W and Mo contents At the moment, we have no clear explanation of this effect. In our opinion, it is probably specific for the stationary testing regime of phosphate-tungstates and phosphate-molybdates i.e. the leached metal atoms react with oxygen diluted in water. This results to the formation of thin oxide films on the surfaces of the ceramics, which prevent further leaching of heavy metals. The surface area covered by the oxide film increases with increasing W and/or Mo contents in ceramics.

To compare radiation stability of phosphate-molybdates and phosphate-tungstates, the irradiation with 167 MeV Xe ions at various fluences ranging from $1\cdot10^{12}$ to $6\cdot10^{13}$ cm$^{-2}$ has been used. The dose dependence of XRD curves registered in Na$_{0.5}$Zr$_2$(PO$_4$)$_{2.5}$(XO$_4$)$_{0.5}$ ceramics is shown in Fig. 17. The results of the XRD investigations have shown the initial Na$_{0.5}$Zr$_2$(PO$_4$)$_{2.5}$(WO$_4$)$_{0.5}$ ceramics (Fig. 17a) to be amorphized under the ion irradiation already at the dose of $3\cdot10^{12}$ cm$^{-2}$. The increasing of the dose resulted in further amorphization and phase decomposition of Na$_{0.5}$Zr$_2$(PO$_4$)$_{2.5}$(WO$_4$)$_{0.5}$ with the formation of the ZrO$_2$ phase. At the irradiation dose of $3\cdot10^{13}$ cm$^{-2}$, no diffraction peaks from the Na$_{0.5}$Zr$_2$(PO$_4$)$_{2.5}$(WO$_4$)$_{0.5}$ phase were observed in the XRD curve, the peaks from the crystalline ZrO$_2$ phase and wide halo from the amorphous component in the specimen remained only.

For the ceramics series of the Na$_{0.5}$Zr$_2$(PO$_4$)$_{2.5}$(MoO$_4$)$_{0.5}$ ceramics (Fig. 17b), the XRD measurement results have shown the ion irradiation to affect the crystallinity of the ceramics weakly. The diffraction peaks are seen clearly even for the irradiation doses up to $6\cdot10^{13}$ cm$^{-2}$. The changes caused by the ion irradiation in the this series concerned the coherent scattering region sizes of the Na$_{0.5}$Zr$_2$(PO$_4$)$_{2.5}$(MoO$_4$)$_{0.5}$ phase slightly. However, the ion irradiation didn't result in a critical degradation of the crystallinity as in the case of the Na$_{0.5}$Zr$_2$(PO$_4$)$_{2.5}$(WO$_4$)$_{0.5}$ ceramics.

### 4. Conclusion

It follows from the analysis of the results obtained that the investigated compounds crystallize in the NASICON-type structure, hexagonal syngony, $R\bar{3}c$ space group for the Na-containing phosphate-molybdates and phosphate-tungstates and $R\bar{3}$ for the Ca-containing ones. The substitution of the phosphate anions (PO$_4$)$^{3-}$ (R(P$^{5+}$) = 0.17 Å) by a larger molybdate (MoO$_4$)$^{2-}$ (R (Mo$^{6+}$) = 0.41 Å) and tungstate (WO$_4$)$^{2-}$ (R(W$^{6+}$) = 0.44 Å) ones leads to an increasing of the unit cell parameters. With a decrease in the population of the extraframe positions in the phosphate-molybdates and phosphate-tungstates investigated, the thermal expansion parameters tend to approach zero.

The ceramics obtained by SPS were featured by high relative densities ($\rho_{rel} > 97.5\%$). The values of microhardness ($H_v$) of the ceramics ranged from 3.60 to 5.64 GPa, the fracture toughness factor ($K_{IC}$) – from 0.86 to 1.62 MPa m$^{1/2}$. According to the X-ray diffraction phase analysis, the structure of the compounds obtained remained unchanged in the course of sintering as well as in the course of the hydrolytic stability tests. The minimum leaching rates achieved were $31\cdot10^{-6}$ g$^{-1}\cdot$cm$^{-2}\cdot$day$^{-1}$ for the Mo compounds and $3.36\cdot10^{-6}$ g$^{-1}\cdot$cm$^{-2}\cdot$day$^{-1}$ for the W ones.

The analysis of the radiation test results demonstrated the destruction of the NZP crystal lattice was less expressed in the Mo-containing samples as compared to the phosphate-tungstates irradiated in similar conditions. The crystal lattice of the W-containing ceramic samples was destructed in the course of irradiation with the fluence $3\cdot10^{13}$ cm$^{-2}$.

**Conflict of interest**.

The authors declare that they have no conflict of interest.

**Acknowledgements**

This work was supported by Russian Science Foundation (Grant No. 21-13-00308).

The TEM study of the powders was carried out on the equipment of the Center Collective Use "Materials Science and Metallurgy" (NUST "MISIS") with the financial support of the Russian Federation represented by the Ministry of Science and Higher Education (grant No. 075-15-2021-696).

The XRD investigations of the specimens after ion irradiation were carried out in Laboratory of Diagnostics of Radiation Defects in Solid State Nanostructures, Institute for Physics of Microstructures, RAS (Project #13) with the financial support of the Russian Federation represented by the Ministry of Science and Higher Education.

Table 1. Crystallographic characteristics of the phosphates Na$_{1-x}$Zr$_2$(PO$_4$)$_{3-x}$(XO$_4$)$_x$ (denoted as "Na") and Ca$_{0.5(1-x)}$Zr$_2$(PO$_4$)$_{3-x}$(XO$_4$)$_x$ (denoted as "Ca")

| X | x | $a$, Å | | $c$, Å | | $V$, Å$^3$ | |
|---|---|---|---|---|---|---|---|
| | | Na | Ca | Na | Ca | Na | Ca |
| - | 0 | 8.799(4) | 8.784(6) | 22.826(7) | 22.736(0) | 1530.6(7) | 1519.4(7) |
| Mo | 0.1 | 8.811(8) | 8.792(0) | 22.856(6) | 22.762(2) | 1536.9(9) | 1523.7(7) |
| | 0.2 | 8.825(5) | 8.801(2) | 22.882(7) | 22.777(7) | 1543.5(3) | 1528.0(2) |
| | 0.3 | 8.833(4) | - | 22.904(9) | - | 1547.8(0) | - |
| | 0.4 | 8.851(9) | - | 22.921(5) | - | 1555.4(2) | - |
| | 0.5 | 8.891(7) | - | 22.897(1) | - | 1567.7(8) | - |
| W | 0.1 | 8.821(6) | 8.813(5) | 22.862(2) | 23.079(8) | 1540.7(7) | 1552.6(0) |
| | 0.2 | 8.827(5) | 8.815(2) | 22.935(1) | 22.949(0) | 1547.7(9) | 1544.3(9) |
| | 0.3 | 8.833(3) | 8.827(4) | 22.989(5) | 22.996(8) | 1553.4(7) | 1551.9(0) |
| | 0.4 | 8.857(1) | - | 23.046(1) | - | 1565.7(0) | - |
| | 0.5 | 8.881(3) | - | 23.053(7) | - | 1574.8(0) | - |

Table 2. Thermal expansion coefficients [in °C$^{-1}$] of phosphates Na$_{1-x}$Zr$_2$(PO$_4$)$_{3-x}$(XO$_4$)$_x$ (denoted as "Na") and Ca$_{0.5(1-x)}$Zr$_2$(PO$_4$)$_{3-x}$(XO$_4$)$_x$ (denoted as "Ca")

| X | x | $\alpha_a \cdot 10^6$ | | $\alpha_c \cdot 10^6$ | | $\alpha_{av} \cdot 10^6$ | | $\beta \cdot 10^6$ | | $\Delta\alpha \cdot 10^6$ | |
|---|---|---|---|---|---|---|---|---|---|---|---|
| | | Na | Ca | Na | Ca | Na | Ca | Na | Ca | Na | Ca |
| - | 0 | -4.20 | -2.96 | 20.37 | 10.60 | 3.99 | 1.56 | 11.82 | 4.60 | 24.56 | 13.56 |
| Mo | 0.1 | -4.43 | -2.73 | 19.82 | 10.10 | 3.66 | 1.55 | 10.90 | 4.71 | 24.25 | 12.83 |
| | 0.2 | -3.63 | -2.61 | 20.50 | 9.57 | 4.41 | 1.45 | 13.14 | 4.30 | 24.12 | 12.18 |
| | 0.3 | -3.62 | - | 20.20 | - | 4.32 | - | 12.75 | - | 23.82 | - |
| | 0.4 | -4.29 | - | 17.49 | - | 2.97 | - | 8.75 | - | 21.79 | - |
| | 0.5 | -2.81 | - | 15.64 | - | 3.34 | - | 9.92 | - | 18.45 | - |
| W | 0.1 | -5.10 | -0.57 | 20.03 | 6.24 | 3.28 | 1.70 | 9.75 | 5.00 | 25.13 | 6.81 |
| | 0.2 | -3.51 | -2.50 | 17.00 | 12.33 | 3.33 | 2.44 | 9.99 | 7.26 | 20.52 | 14.83 |
| | 0.3 | -2.94 | -2.26 | 15.09 | 10.80 | 3.07 | 2.09 | 9.23 | 6.29 | 18.04 | 13.06 |
| | 0.4 | -2.71 | - | 11.85 | - | 2.14 | - | 6.45 | - | 14.56 | - |
| | 0.5 | -3.71 | - | 11.65 | - | 1.41 | - | 4.23 | - | 15.36 | - |

Table 3. Characteristics of ceramic samples of phosphates $Na_{1-x}Zr_2(PO_4)_{3-x}(XO_4)_x$

| X | x | $V_h$, °C/min | $T_s$, °C | $t_{sps}$, min | ρ, % | $H_v$, GPa | $K_{IC}$, MPa·m$^{1/2}$ |
|---|---|---|---|---|---|---|---|
| Mo | 0.1 | 50 | 920 | 12.58 | 97.55 | 3.6 | 1.0 |
|    | 0.2 |    | 932 | 13.08 | 98.42 | - | - |
|    | 0.3 |    | 1110 | 16.50 | 100.07 | 5.3 | 0.9 |
|    | 0.4 |    | 975 | 13.58 | 100.06 | - | - |
|    | 0.5 |    | 861 | 11.50 | 101.66 | 5.3 | 1.4 |
| W  | 0.1 | 50 | 1190 | 18.08 | 97.70 | 5.3 | 1.3 |
|    | 0.2 |    | 1100 | 16.83 | 98.75 | 5.6 | 1.6 |
|    | 0.3 |    | 1110 | 17.58 | 98.59 | 5.6 | 1.5 |
|    | 0.4 |    | 1065 | 15.67 | 100.20 | 5.5 | 1.1 |
|    | 0.5 |    | 1065 | 15.08 | 100.70 | 5.6 | - |

Table 4. Normalized weight losses (NL) and leaching rates (R) of Mo and W from Na$_{1-x}$Zr$_2$(PO$_4$)$_{3-x}$(XO$_4$)$_x$ ceramics

| x | t, days | m·10$^4$, g | | NL·10$^2$, g/cm$^2$ | | R·10$^5$, g/(cm$^2$·days) | |
|---|---|---|---|---|---|---|---|
| | | Mo | W | Mo | W | Mo | W |
| 0.4 | 1 | 5.917 | 8.333 | 0.453 | 0.352 | 60.000 | 12.300 |
| | 3 | 1.417 | 0.417 | 0.561 | 0.369 | 23.583 | 4.289 |
| | 7 | 0.958 | 0.275 | 0.634 | 0.381 | 11.477 | 1.903 |
| | 10 | 0.375 | 0.133 | 0.663 | 0.387 | 8.475 | 1.352 |
| | 14 | 0.375 | 0.125 | 0.692 | 0.392 | 6.367 | 0.979 |
| | 21 | 0.458 | 0.167 | 0.727 | 0.399 | 4.511 | 0.664 |
| | 28 | 0.375 | 0.125 | 0.755 | 0.404 | 3.532 | 0.504 |
| 0.5 | 1 | 7.250 | 7.500 | 0.427 | 0.246 | 55.600 | 8.200 |
| | 3 | 1.833 | 0.392 | 0.535 | 0.259 | 21.591 | 2.859 |
| | 7 | 0.958 | 0.208 | 0.591 | 0.266 | 10.410 | 1.269 |
| | 10 | 0.433 | 0.125 | 0.617 | 0.270 | 7.657 | 0.901 |
| | 14 | 0.400 | 0.108 | 0.640 | 0.274 | 5.731 | 0.653 |
| | 21 | 0.442 | 0.167 | 0.666 | 0.279 | 4.042 | 0.442 |
| | 28 | 0.408 | 0.167 | 0.690 | 0.285 | 3.156 | 0.336 |

**List of Figures**

Figure 1. Regular sintering diagram "temperature – pressure – time" for SPS. $Na_{0.7}Zr_2(PO_4)_{2.7}(MpO_4)_{0.3}$ compound

Figure 2. SEM images of the agglomerates (a, c, e) and of the synthesized powders (b, d, f): $Ca_{0.4}Zr_2(MoO_4)_{0.1}(PO_4)_{2.9}$ (a, b), $Na_{0.5}Zr_2(MoO_4)_{0.5}(PO_4)_2$ (c, d), $Na_{0.5}Zr_2(WO_4)_{0.5}(PO_4)_2$ (e, f)

Figure 3. TEM images of the powders $Ca_{1-x}Zr_2(PO_4)_{3-x}(MoO_4)_x$ with $x = 0.1$ (a, b), $x = 0.3$ (c, d), $x = 0.4$ (e, f)

Figure 4. TEM images of the powders $Na_{0.5}Zr_2(PO_4)_{2.5}(XO_4)_{0.5}$: $X = Mo$ (a, b, e), $X = W$ (c, d, f)

Figure 5. XRD data. Phosphates $Na_{1-x}Zr_2(PO_4)_{3-x}(XO_4)_x$ compounds, $X = Mo$ (1), $W$ (2), $x = 0$ (a), 0.1 (b), 0.2 (c), 0.3 (d), 0.4 (e), 0.5 (f)

Figure 6. XRD data. Phosphates $Ca_{0.5(1-x)}Zr_2(PO_4)_{3-x}(XO_4)_x$ compounds, $X = Mo$ (1), $W$ (2), $x = 0$ (a), 0.1 (b), 0.2 (c), 0.3 (d), 0.4 (e), 0.5 (f)

Figure 7. Dependencies of the CSR sizes ($D_{XRD}$) on the Mo and W contents in the $Na_{1-x}Zr_2(PO_4)_{3-x}(XO_4)_x$ powders

Figure 8. Dependencies of the unit cell parameters $a$ (1) and $c$ (2) and of the unit cell volume $V$ (3) on the composition ($x$) of phosphates $Na_{1-x}Zr_2(PO_4)_{3-x}(XO_4)_x$ (I) and $Ca_{0.5(1-x)}Zr_2(PO_4)_{3-x}(XO_4)_x$ (II), $X = Mo$ (a), $W$ (b)

Figure 9. Temperature dependencies of the unit cell parameters $a$ (1) and $c$ (2) and of the unit cell volume $V$ (3) of phosphates of $Na_{1-x}Zr_2(PO_4)_{3-x}(XO_4)_x$, X = Mo (I), W (II); $x$ = 0 (a), 0.1 (b), 0.2 (c), 0.3 (d), 0.4 (e), 0.5 (f)

Figure 10. Temperature dependencies of the unit cell parameters $a$ (1) and $c$ (2) and of the unit cell volume $V$ (3) of phosphates $Ca_{0.5(1-x)}Zr_2(PO_4)_{3-x}(XO_4)_x$, X = Mo (I), W (II); $x$ = 0 (a), 0.1 (b), 0.2 (c), 0.3 (d)

Figure 11. Dependencies of the thermal expansion parameters ($\alpha_a$, $\alpha_c$, $\alpha_{av}$, $\beta$, and $\Delta\alpha$) on the compositions ($x$) of phosphates $Na_{1-x}Zr_2(PO_4)_{3-x}(XO_4)_x$, X = Mo (1), W (2)

Figure 12. Dependencies of the thermal expansion parameters ($\alpha_a$, $\alpha_c$, $\alpha_{av}$, $\beta$, and $\Delta\alpha$) on the compositions ($x$) of phosphates $Ca_{0.5(1-x)}Zr_2(PO_4)_{3-x}(XO_4)_x$, X = Mo (1), W (2)

Figure 13. SPS diagrams L(T) (a, b) and S(T) (c, d) for the $Na_{1-x}Zr_2(PO_4)_{3-x}(XO_4)_x$ ceramics, X = Mo (a, c), W (d, d)

Figure 14. Microstructure of $Na_{1-x}Zr_2(PO_4)_{3-x}(MoO_4)_x$ ceramics with $x$ = 0.1 (a, b) and $x$ = 0.5 (c, d). SEM images of the fractures of sintered ceramic samples

Figure 15. Microstructure of $Na_{1-x}Zr_2(PO_4)_{3-x}(WO_4)_x$ ceramics with $x$ = 0.1 (a, b) and $x$ = 0.5 (c, d). SEM images of the fractures of sintered ceramic samples

Figure 16. Time dependences of the normalized weight loss NL (a) and of the release rates for particular components per unit surface area R (b) for the $Na_{1-x}Zr_2(PO_4)_{3-x}(XO_4)_x$ ceramics

Figure 17. XRD curves of the ceramic samples of phosphate-tungstates (a) and phosphate-molybdates (b) with $x$ = 0.5 after irradiation with fluences (cm$^{-2}$): the initial one and irradiated by different Xe ion doses (in cm$^{-2}$): Fig.17a: W1 – 3·10$^{12}$; W2 – 10$^{13}$; W3 – 3·10$^{13}$; Fig.17b: M1 – 10$^{12}$; M2 – 3·10$^{12}$; M3 – 6·10$^{12}$; M4 – 8·10$^{12}$; M5 – 10$^{13}$; M6 – 3·10$^{13}$; M7 – 6·10$^{13}$

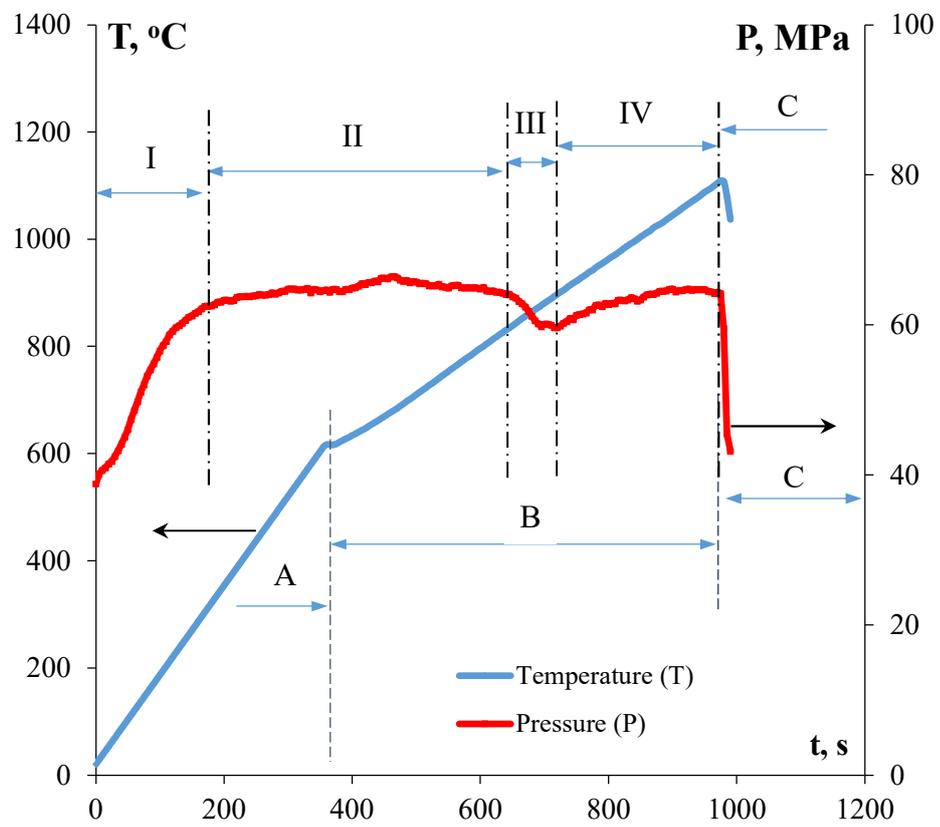

Figure 1

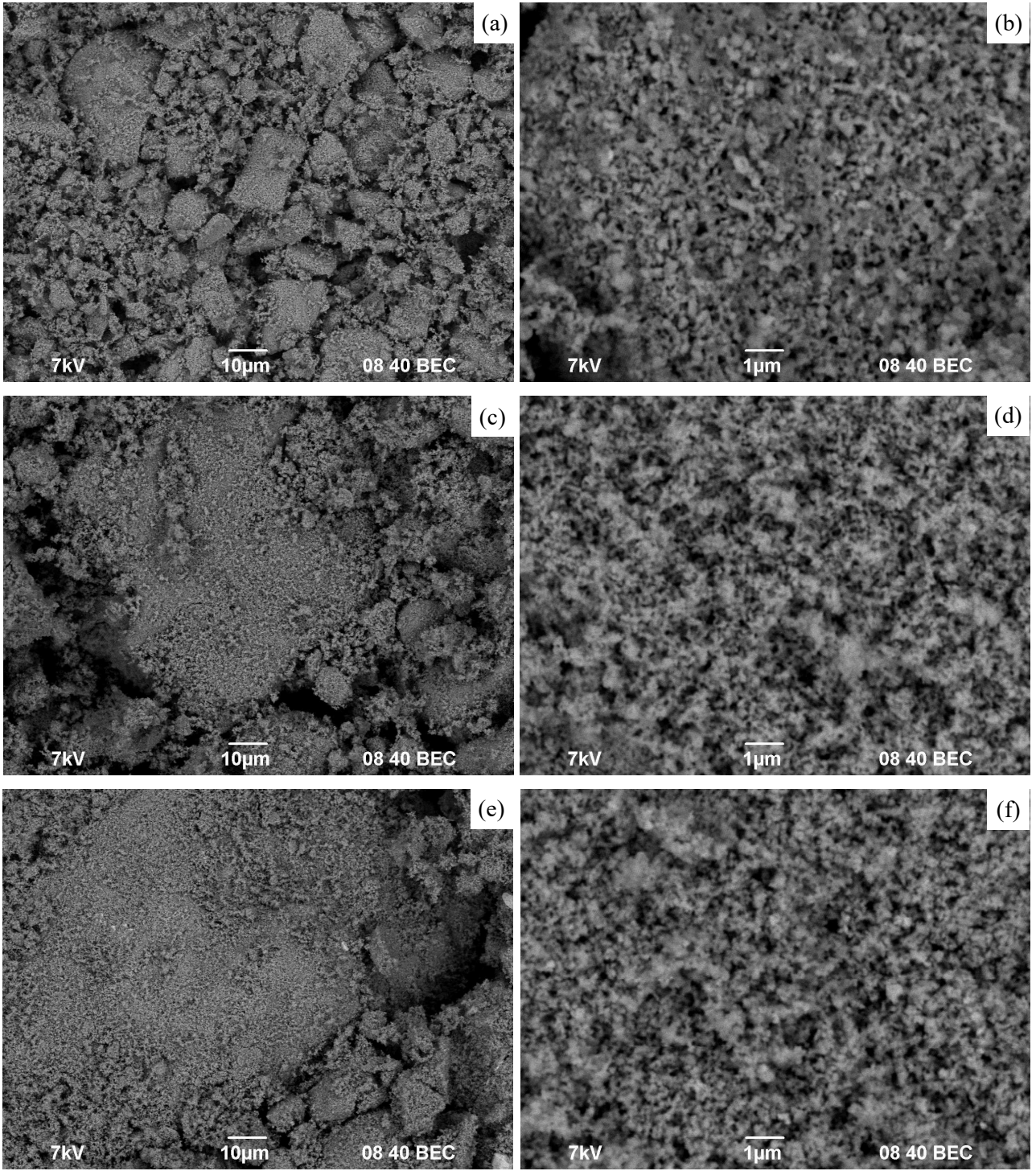

Figure 2

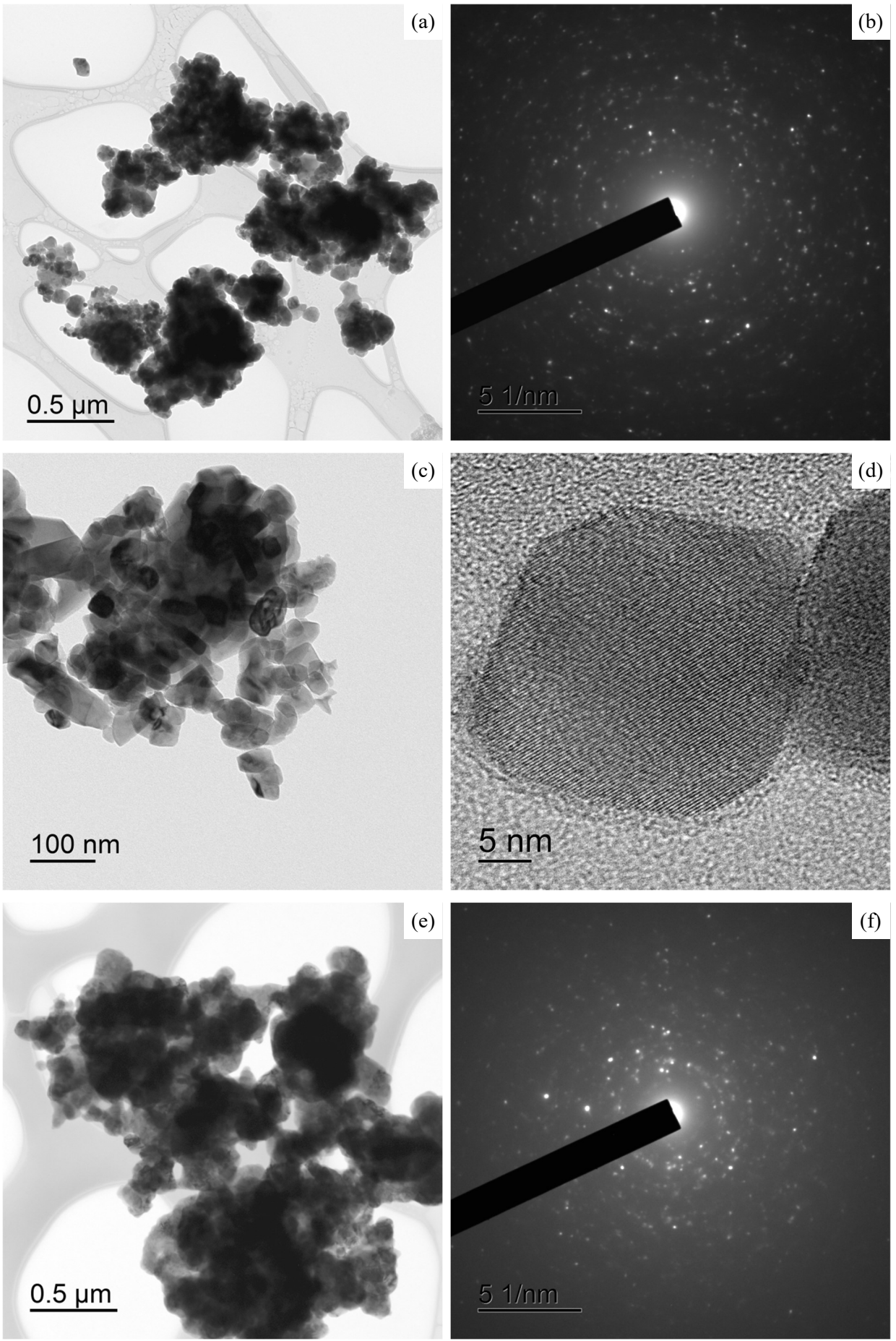

Figure 3

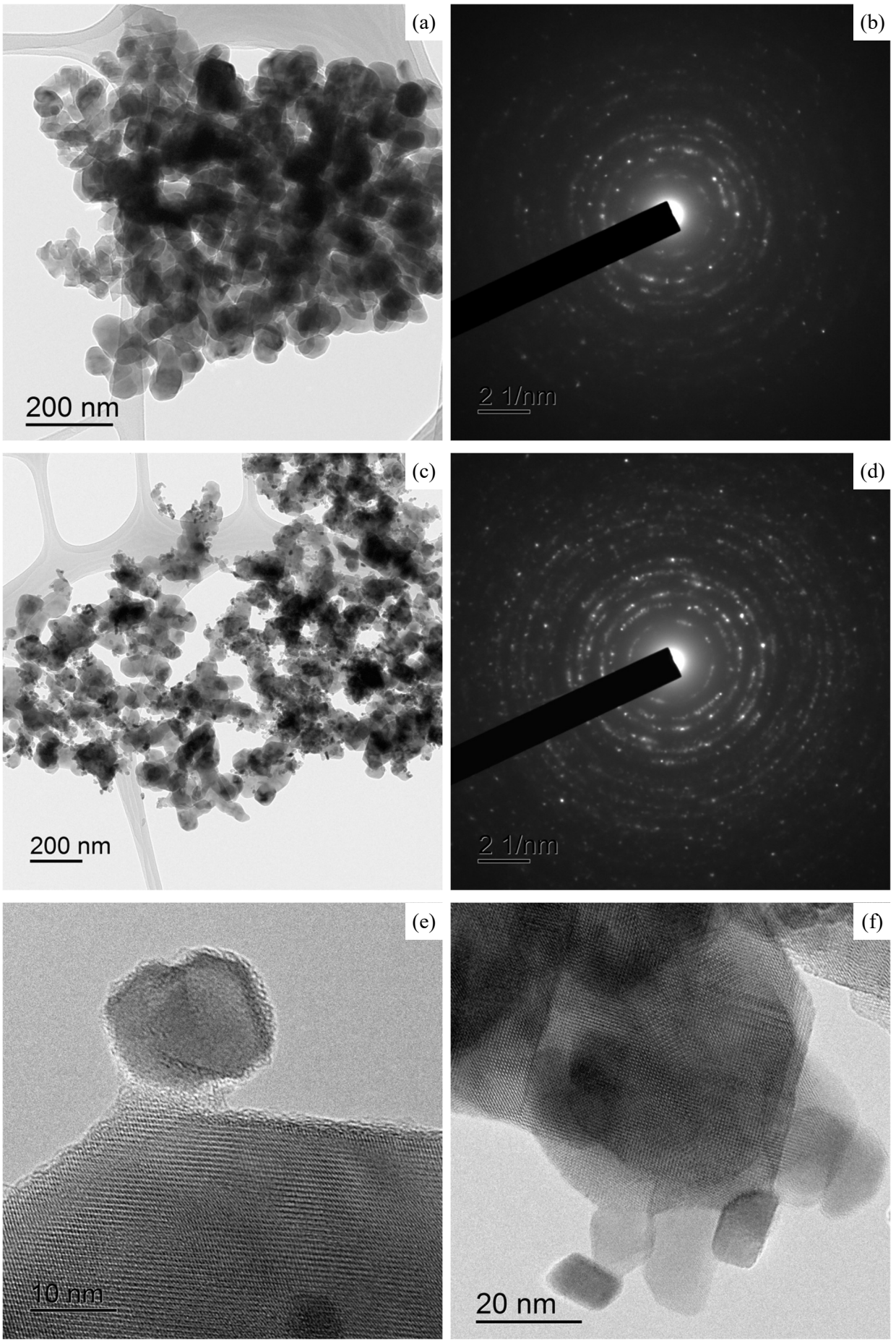

Figure 4

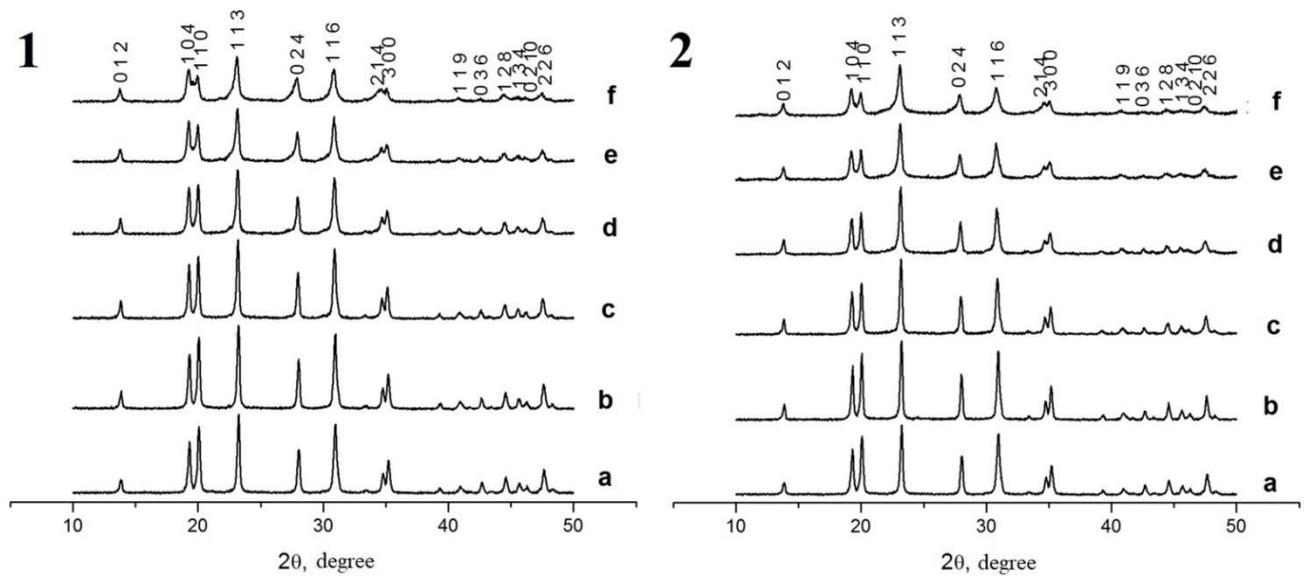

Figure 5

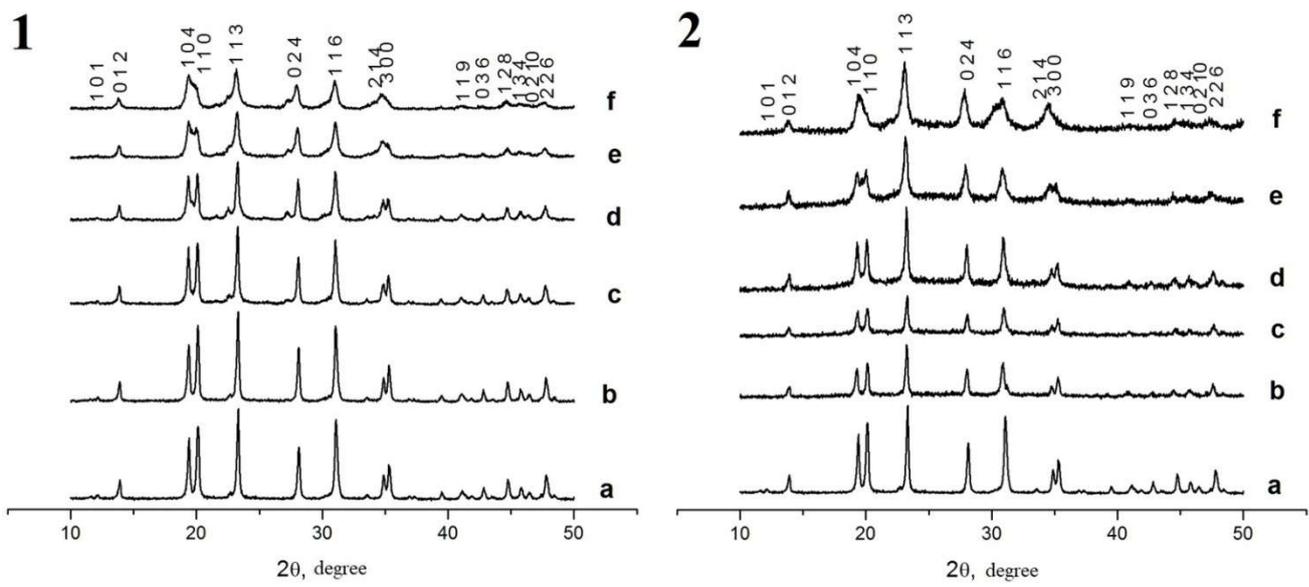

Figure 6

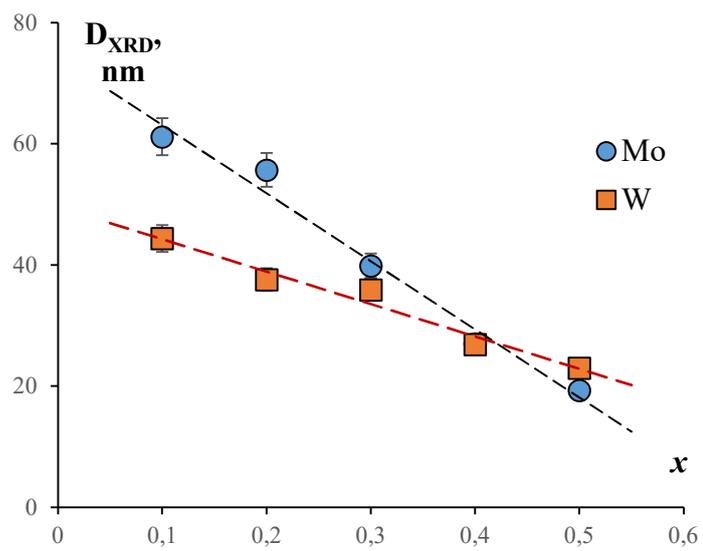

Figure 7

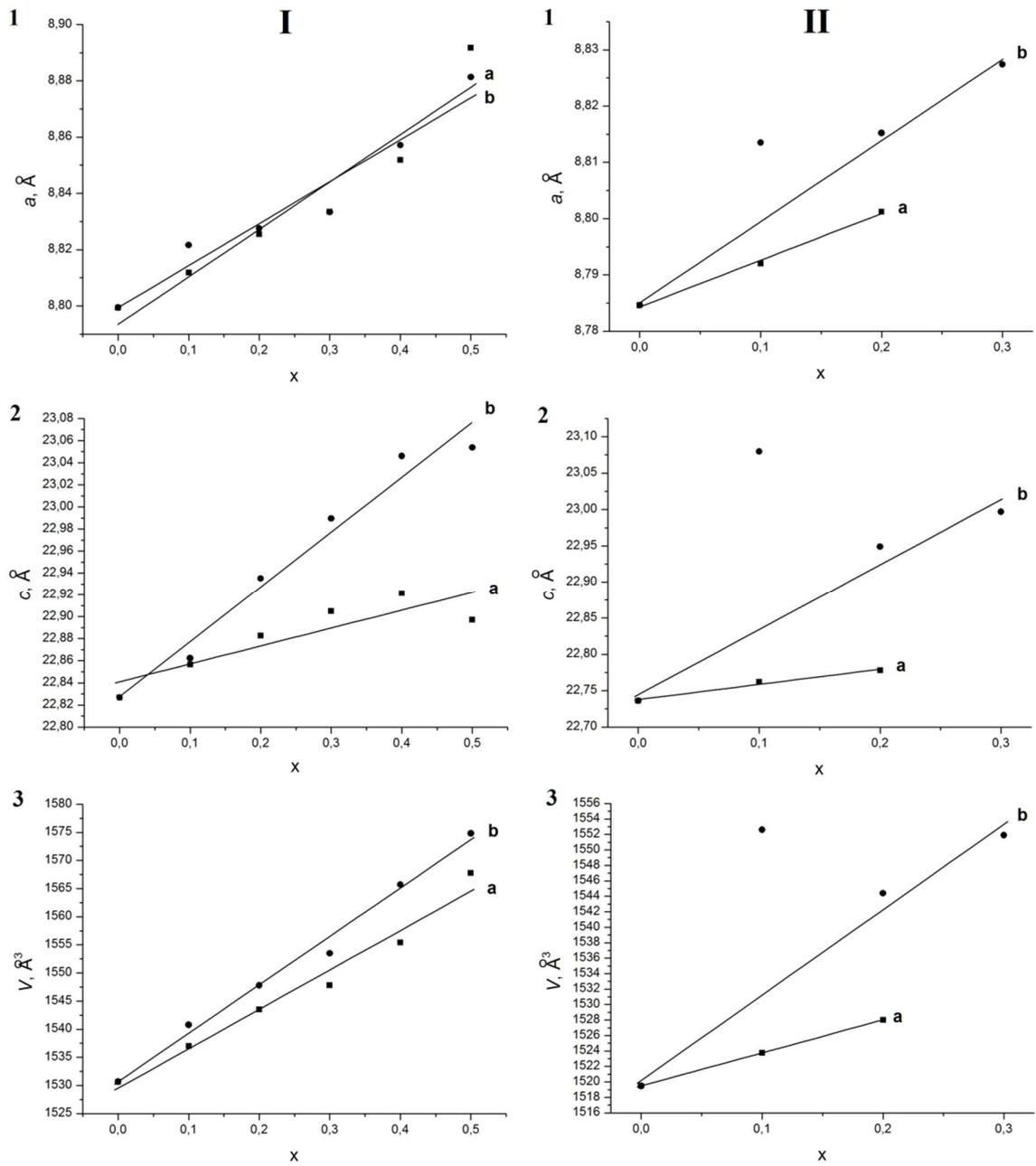

Figure 8

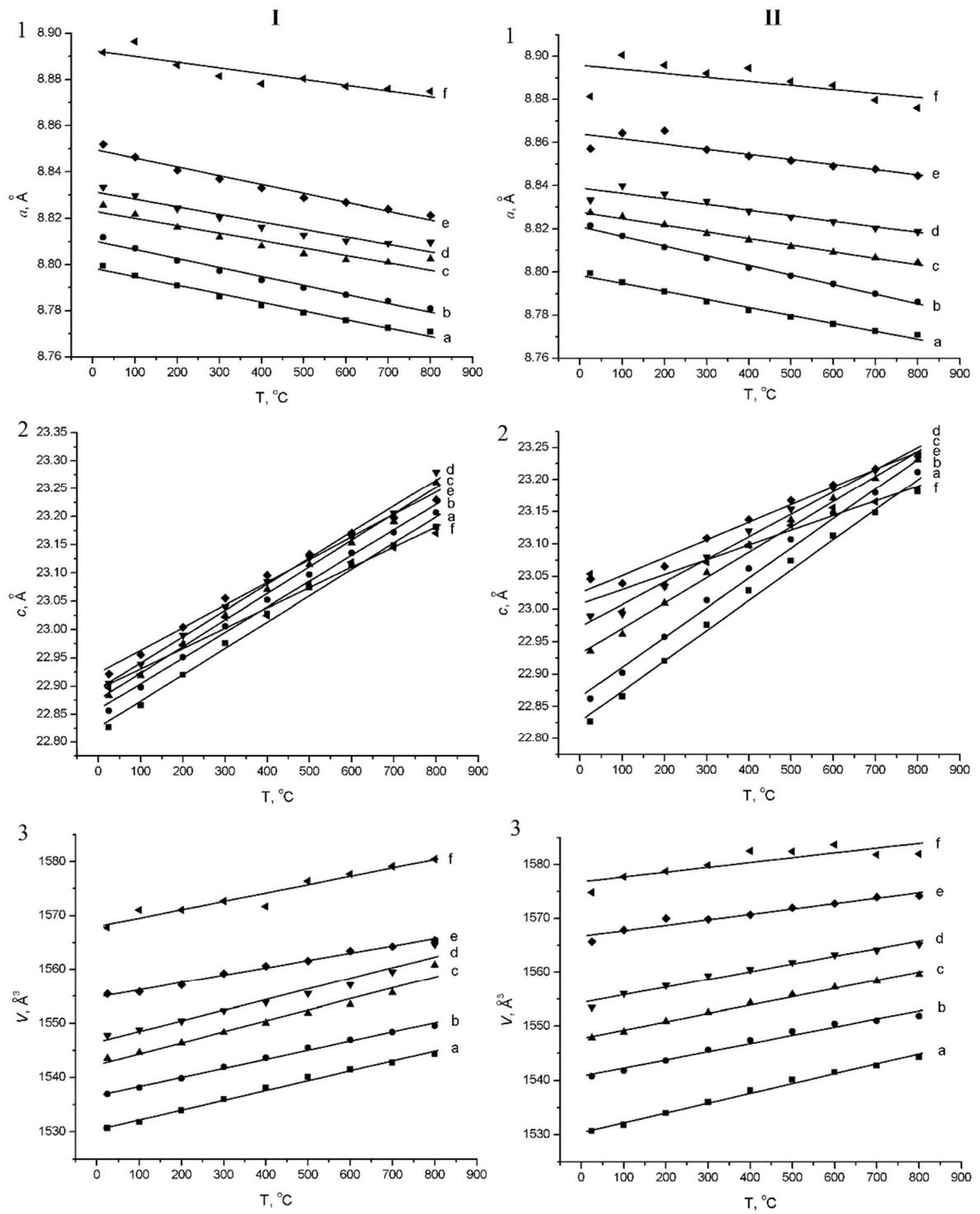

Figure 9

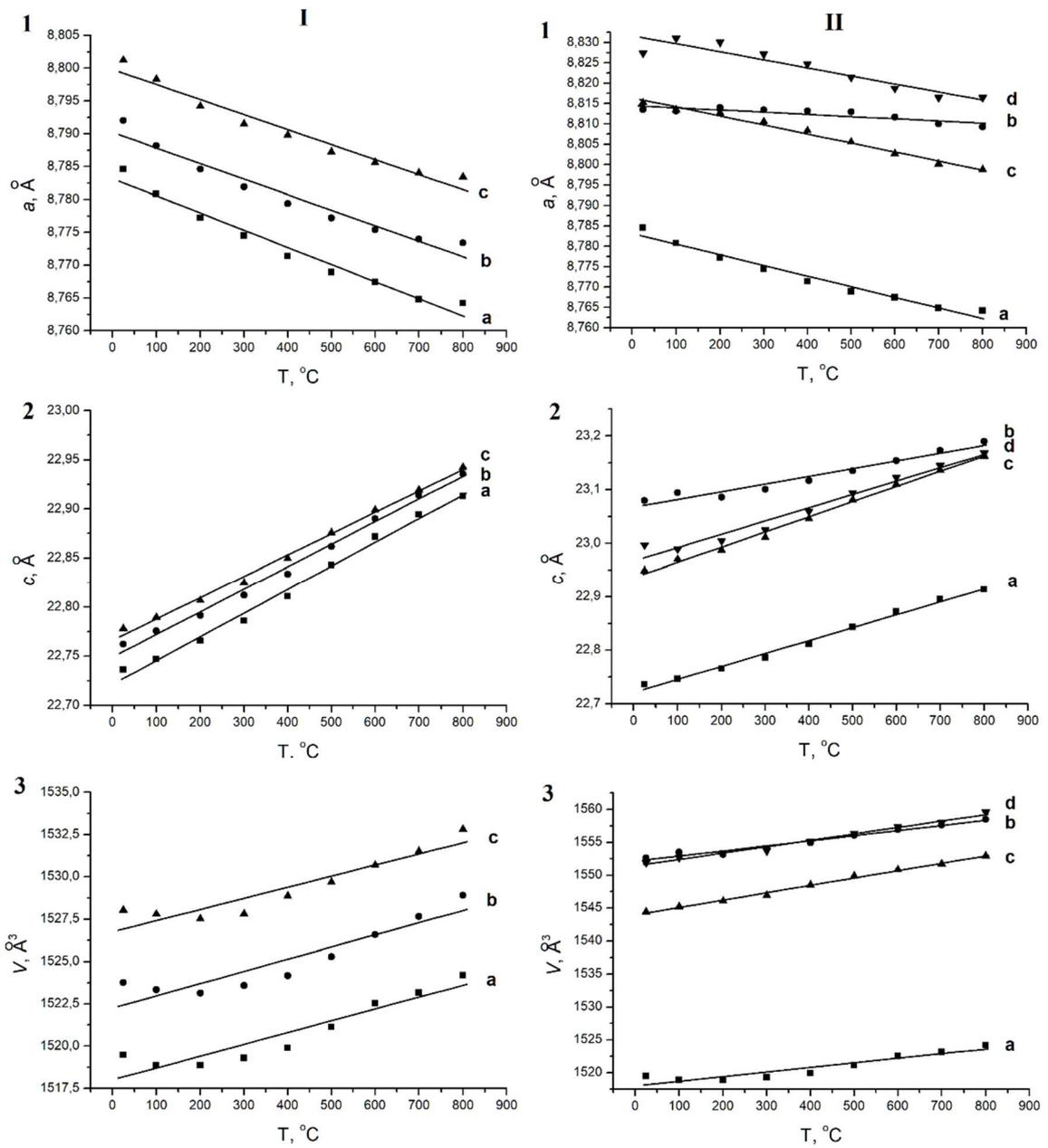

Figure 10

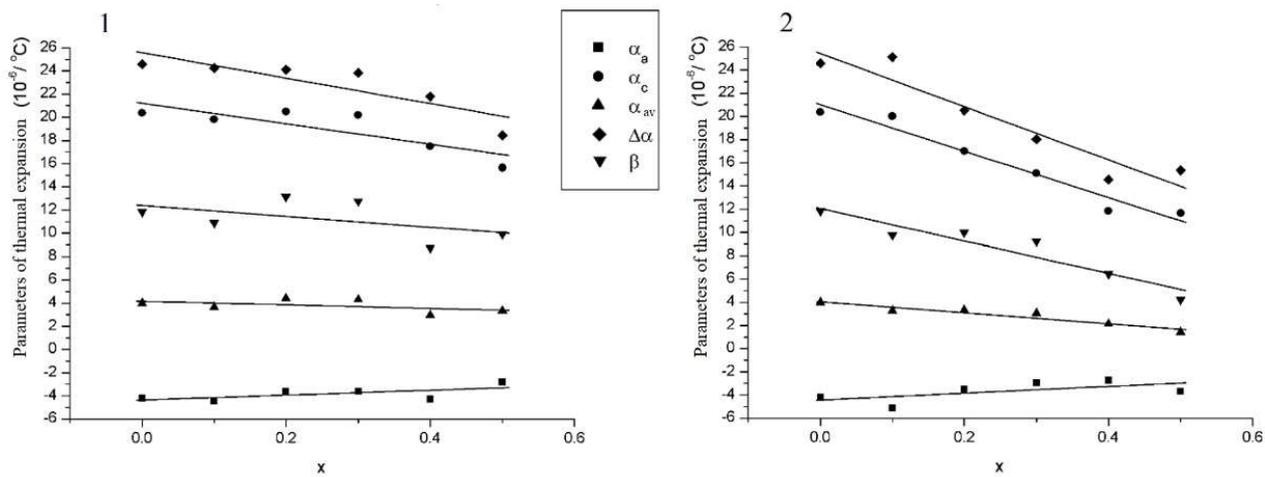

Figure 11

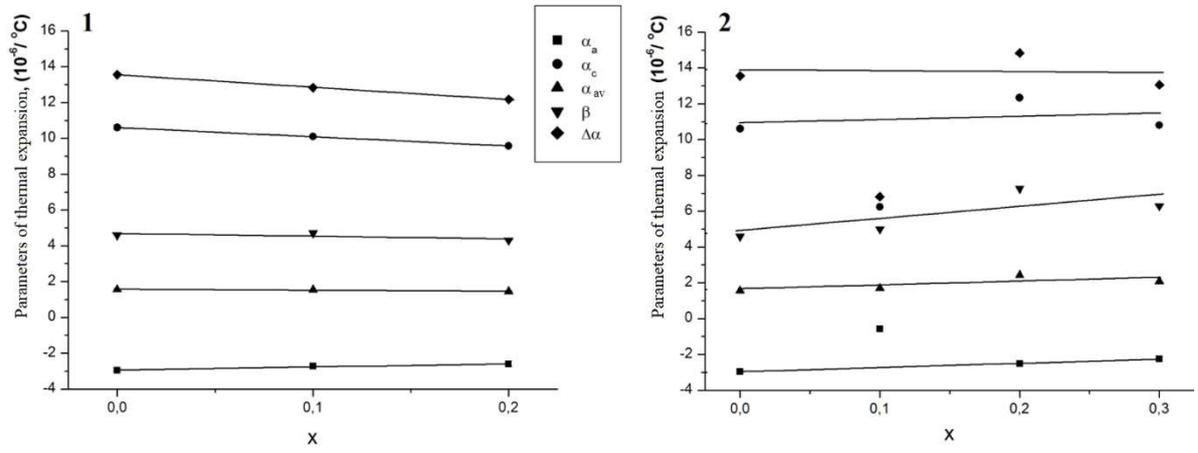

Figure 12

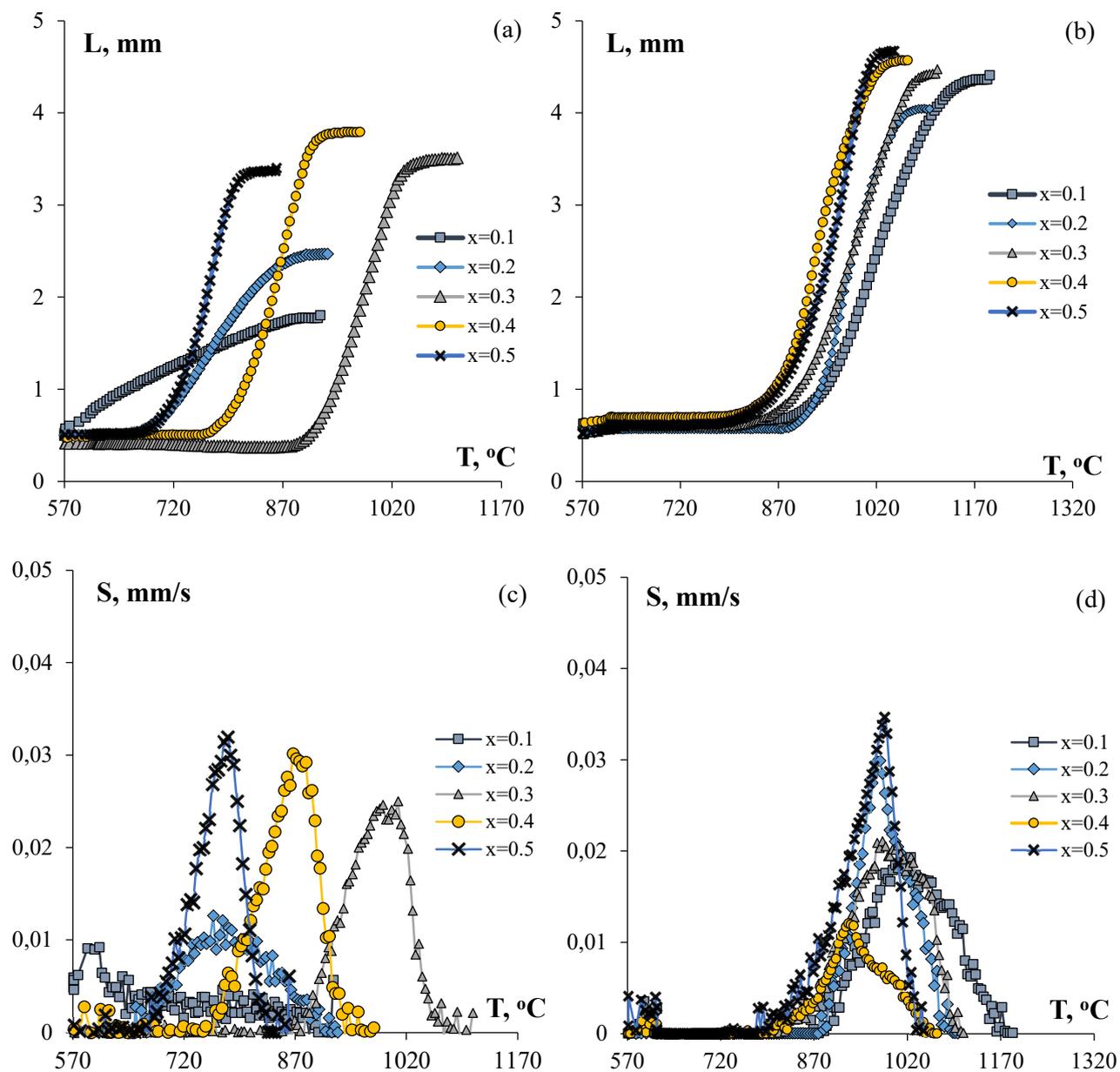

Figure 13

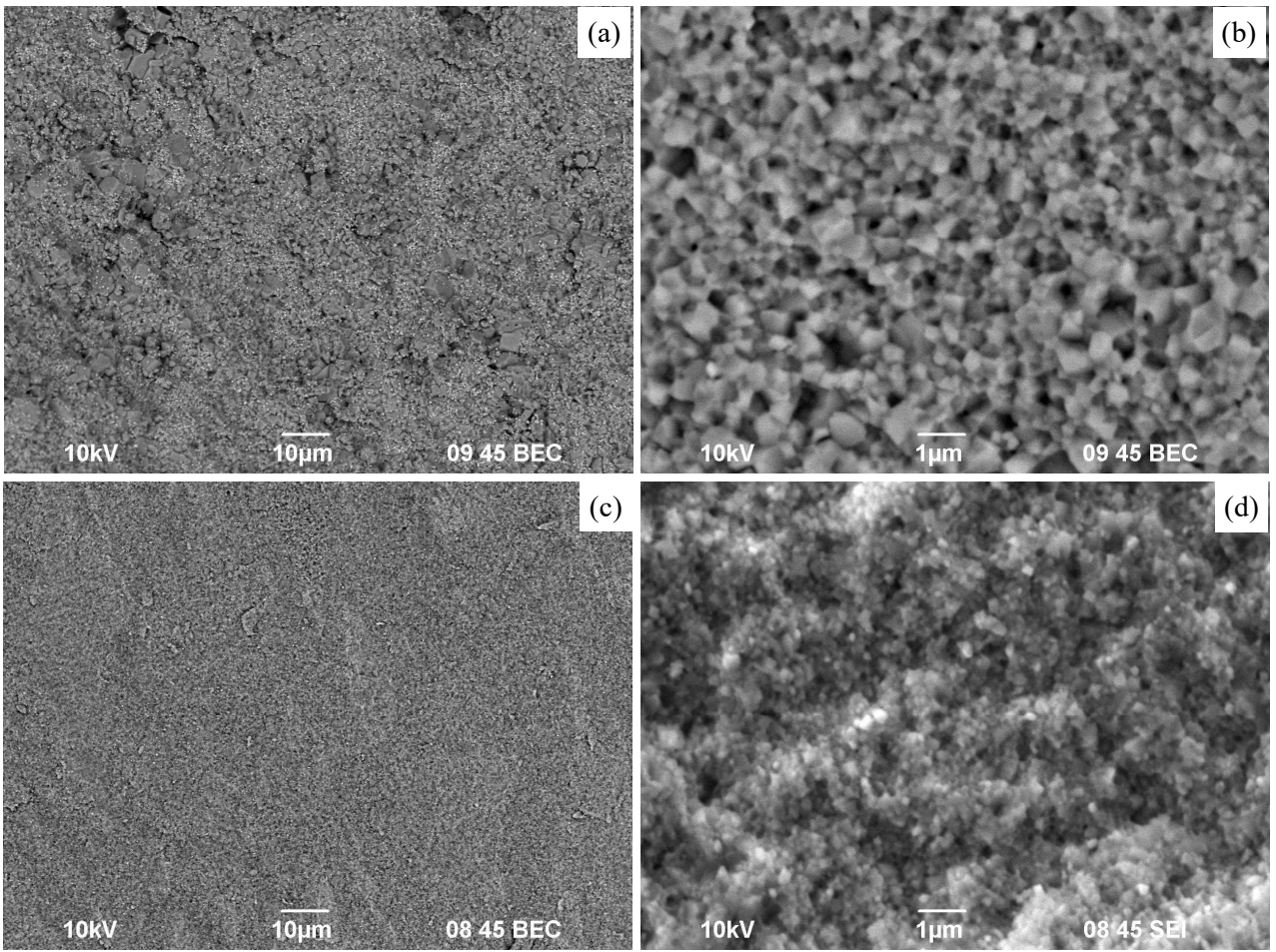

Figure 14

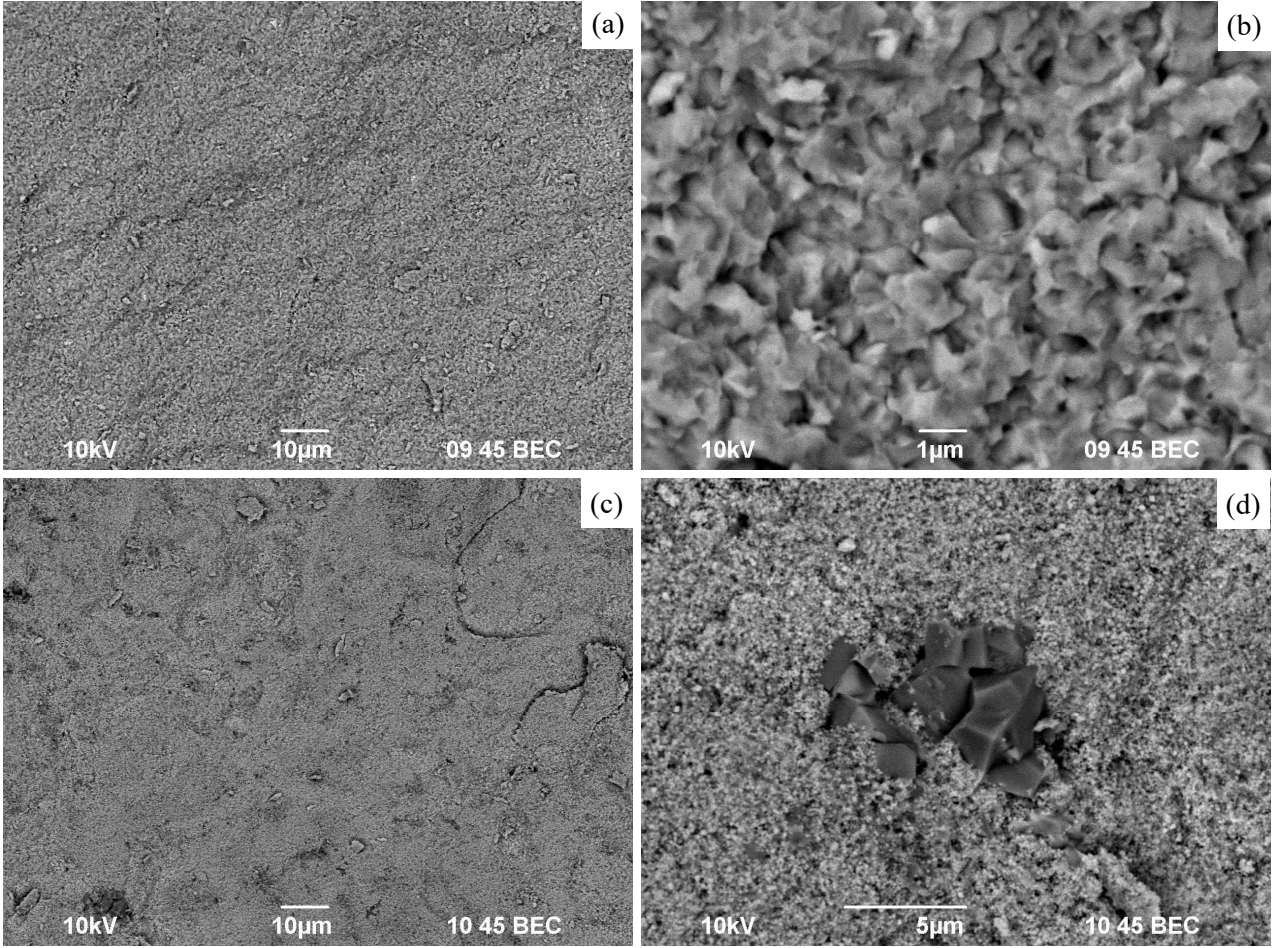

Figure 15

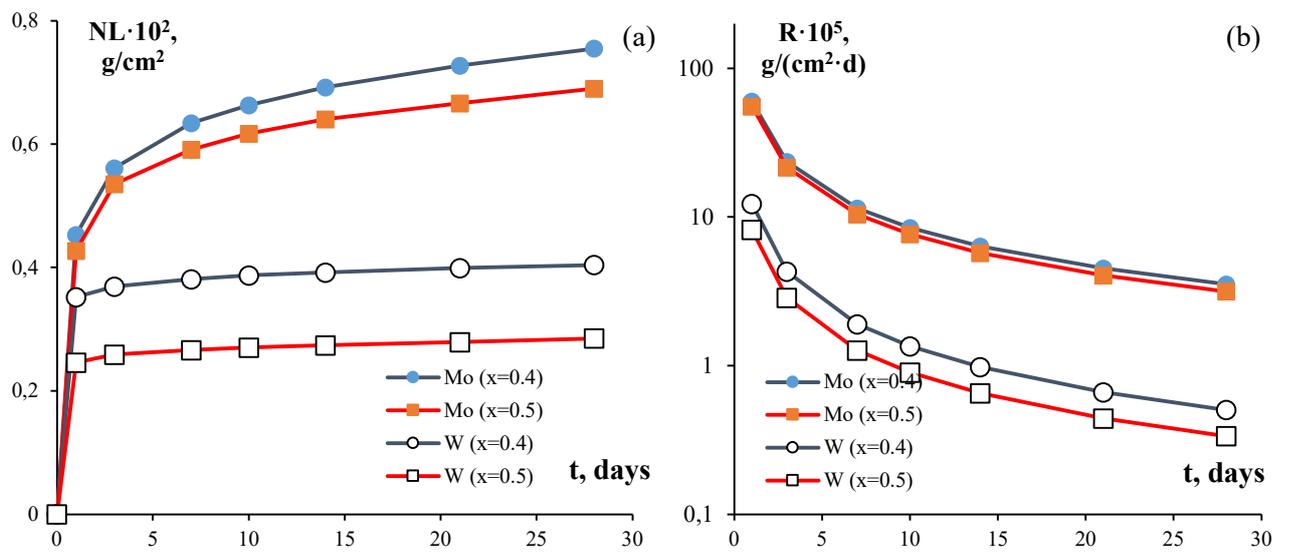

Figure 16

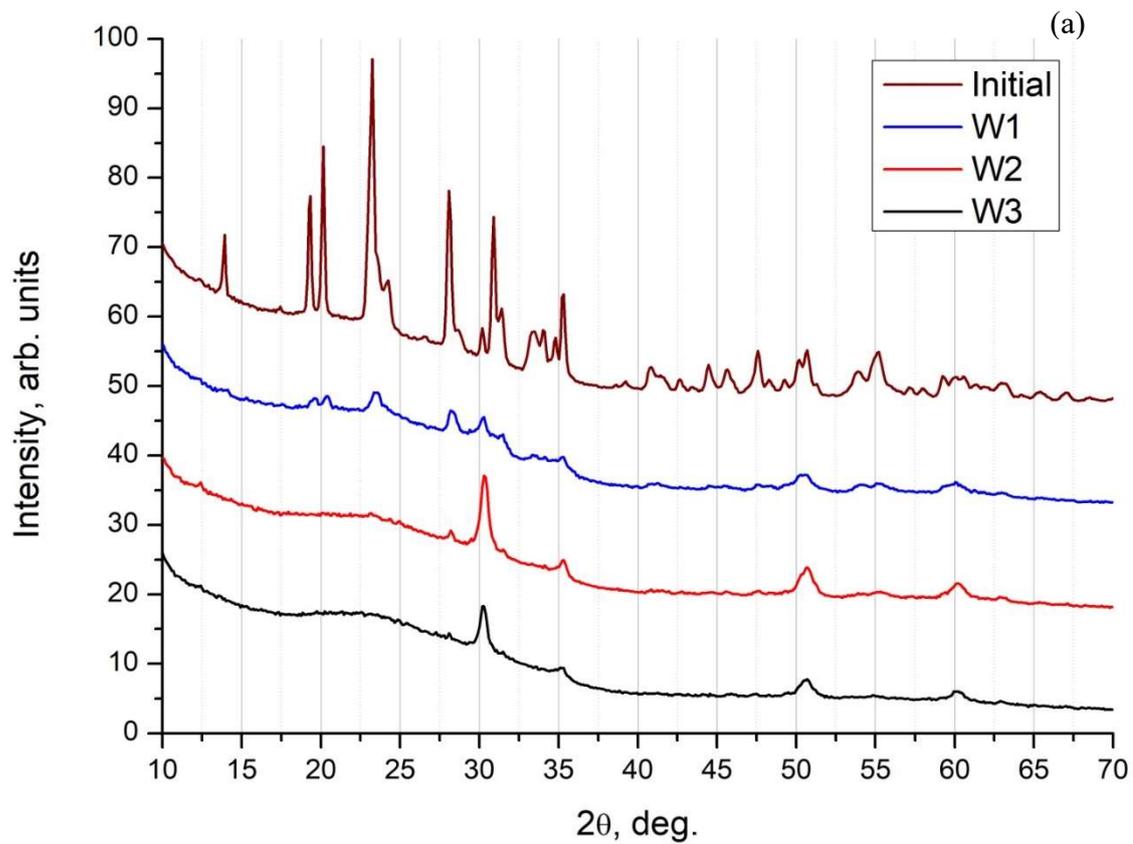

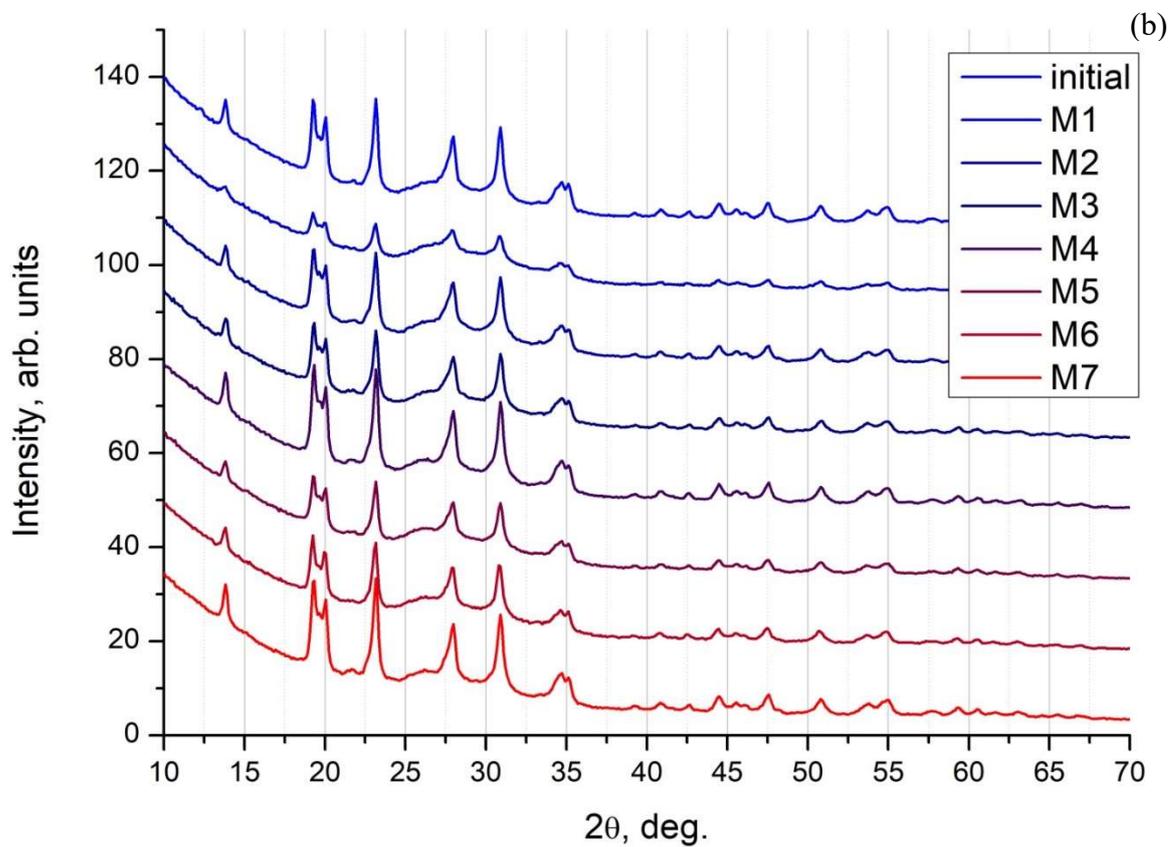

Figure 17